\documentclass{article}
\usepackage{amssymb}
\usepackage{amsfonts}
\usepackage{amsmath}

\input{tcilatex}
\begin{document}

\title{Atomic radiative corrections without QED: role of the zero-point field%
}
\author{Ana Mar\'{\i}a Cetto*, \and Luis de la Pe\~{n}a, \and Andrea Vald%
\'{e}s-Hern\'{a}ndez \\
Instituto de F\'{\i}sica, Universidad Nacional Aut\'{o}noma de M\'{e}xico,\\
Apartado postal 20-364, 01000 M\'{e}xico, DF, Mexico\\
E-mail addresses: ana, luis, andreavh@fisica.unam.mx\\
Corresponding author}
\maketitle

\begin{abstract}
We derive the atomic radiative corrections predicted by \textsc{qed} using
an alternative approach that offers the advantage of physical clarity and
transparency. The element that gives rise to these corrections is the
fluctuating zero-point radiation field (\textsc{zpf}) of average energy $%
\hbar \omega /2$ per mode, which ---in contrast with \textsc{qed}--- is
taken here as a primordial real entity in permanent interaction with matter
and responsible for its quantization. After briefly recalling how quantum
mechanics itself emerges as a result of the balance between the zpf and
radiation reaction, the most important higher-order effects of the radiative
terms on the atom are studied. The nonrelativistic \textsc{qed} formulas for
the lifetimes and the Lamb shift, as well as the corrections to the latter
due to external factors that modify the vacuum field, are thus obtained in a
self-consistent approach and without the need to resort to second
quantization to the present order of approximation.

Keywords: Radiative corrections, Atomic Lamb shift, Atomic lifetimes,
Zero-point field
\end{abstract}

\section{Introduction}

The random zero-point radiation field (\textsc{zpf}) of mean energy $\hbar
\omega /2$ per normal mode, taken as a real field, has been shown in a
series of recent papers \cite{PeVaCe09}-\cite{CePeVa12} to be responsible
for the basic quantum properties of matter. In particular, the usual quantum
description, as afforded, e.g. by the Schr\"{o}dinger equation, is obtained
as a result of reducing the original phase-space description of the entire
particle-\textsc{zpf} system to the configuration space of the particle in
the time-asymptotic limit, in which an energy balance is reached between
radiation reaction and the \textsc{zpf} and higher-order effects of the
radiative terms can be neglected.

This paper is concerned with the main effects on the atom of the previously
neglected radiative terms, calculated to lowest order in $\alpha
=e^{2}/\hbar c$. After briefly reviewing some consequences of the
energy-balance condition and the role of the \textsc{zpf} in fixing the
atomic stationary states, an analysis of the dynamics in the absence of
energy balance is made, leading to formulas for the radiative lifetimes of
excited states. Further, a calculation of the contribution of the radiative
terms to the average energy gives the nonrelativistic formula for the Lamb
shift. Finally, a modification of the background field through the presence
of an external field or material objects is shown to produce in general a
change in the radiative lifetimes and a shift of the atomic energy levels.

The nonrelativistic, spinless, electric dipole approximation is made
throughout the paper. The correct results (i.e. those predicted by
non-relativistic quantum electrodynamics, \textsc{qed}) are obtained in all
cases. With these results we demonstrate that the theory of stochastic
electrodynamics in its present form \cite{PeVaCe09}-\cite{CePeVa12} takes us 
\emph{beyond }quantum mechanics, to the realm of \textsc{qed}. Though most
of the results for the radiative corrections derived in the present work are
well known, their connection with the condition of energy balance between
the \textsc{zpf} and radiation reaction is not. An interesting point is that
in each case explicit formulas for the effects discussed are obtained, along
with a clear physical picture of their meaning

The theory used here should be clearly distinguished from what is called
semiclassical theory (e.g. \cite{ScZu97}{\Large )}. The \textsc{sed}
approach is \textit{not} an attempt to replace quantum physics with a
classical (or semiclassical) theory; quite the contrary, it is an endeavour
intended to give deep physical support to quantum theory by answering
fundamental questions as, e.g., on the physical mechanism that leads to the
quantization and the stability of the atom, or the physical cause and nature
of the quantum fluctuations. In this approach the \textsc{zpf} is seen to
play a crucial role for atomic stability through the energy-balance
condition; there is no quantization in the absence of this field. The
already quantized atom continues to interact with the \textsc{zpf}, which
leads to the radiative corrections here studied. Although in the quantum
regime also the field satisfies quantum rules, as discussed in \cite{PeCe07}%
, \cite{PeVaCe08}, these are not explicitly needed for the present purposes.
Quantized matter under the action of the \textsc{zpf} is sufficient to
obtain the nonrelativistic radiative corrections to lowest order in $\alpha $%
.

\section{The quantum regime}

\subsection{Radiationless approximation\label{QM}}

For clarity in the exposition, let us recall in this section the main steps
leading to the Schr\"{o}dinger equation on the basis of the existence of the 
\textsc{zpf}. For details see \cite{PeVaCeFr11}, \cite{CePeVa12}.

The motion of the particle is governed in the nonrelativistic limit by the
equations (we use one-dimensional notation wherever possible, for simplicity)%
\begin{equation}
\dot{x}=p/m,\text{ \quad }\dot{p}=f(x)+m\tau \dddot{x}+eE(t),  \label{eqsmot}
\end{equation}%
where $f(x)$ is the external force, $E(t)$ is the electric component of the
random \textsc{zpf} in the long-wavelength approximation and $m\tau \overset{%
...}{x}$ is the radiation reaction force in the Abraham-Lorentz
approximation, with $\tau =2e^{2}/3mc^{3}$ ($\approx 10^{-23}$ s for the
electron). The density $R$ of points in the particle's phase space is
determined by%
\begin{equation}
\frac{\partial }{\partial t}R+\frac{\partial }{\partial x}(\dot{x}R)+\frac{%
\partial }{\partial p}\left( f(x)+m\tau \dddot{x}\right) R=-\frac{\partial }{%
\partial p}E(t)R.  \label{contin}
\end{equation}%
Averaging over the realizations of the field one obtains for the (mean)
density in the phase space of the particle, $\overline{R(x,p,t)}^{E}\equiv $ 
$Q(x,p,t),$ the generalized Fokker-Planck equation%
\begin{equation}
\frac{\partial }{\partial t}Q+\hat{L}Q=e^{2}\frac{\partial }{\partial p}\hat{%
D}(t)Q,  \label{FPE}
\end{equation}%
with $\hat{L}$ the Liouville operator%
\begin{equation}
\hat{L}=\frac{1}{m}\frac{\partial }{\partial x}p+\frac{\partial }{\partial p}%
(f+m\tau \dddot{x}),  \label{Liou}
\end{equation}%
and $\hat{D}$ the diffusion operator, given to first order in $e^{2}$ by 
\begin{equation}
e^{2}\hat{D}(t)Q=e^{2}\int^{t}dt^{\prime }\overline{E(t)E(t^{\prime })}%
^{E}e^{-\hat{L}(t-t^{\prime })}\frac{\partial Q}{\partial p}.  \label{diff}
\end{equation}%
The correlation of the electric field components is related with the
spectral energy density of the field through 
\begin{equation}
\overline{E(t)E(t^{\prime })}^{E}=(4\pi /3)\int\limits_{0}^{\infty }\rho
(\omega )\cos \omega (t-t^{\prime })d\omega .  \label{corrE}
\end{equation}%
For the \textsc{zpf} (the field at temperature $T=0$) the spectral energy
density is given by%
\begin{equation}
\rho (\omega ,T)_{T=0}=\rho _{0}(\omega )=\frac{\hbar \omega ^{3}}{2\pi
^{2}c^{3}}.  \label{rho0}
\end{equation}%
\qquad

To make the transition from the phase-space equation (\ref{FPE}) to a
description in configuration space, the characteristic function $\widetilde{Q%
}(x,z,t)=$ $\int Q(x,p,t)e^{ipz}dp$ is introduced, so that the marginal
probability density is $\rho (x,t)=\int Q(x,p,t)dp=\widetilde{Q}(x,0,t).$ By
expanding the Fourier transform of Eq. (\ref{FPE}) into a power series
around $z=0$ and separating the coefficients of $z^{k}$ ($k=0,1,2,\ldots $),
a hierarchy of coupled equations for moments of $p$ of increasing order is
obtained. The first two are the continuity equation and the equation for the
transfer of momentum, which with the help of the change of variables 
\begin{equation}
z_{\pm }=x\pm \eta z  \label{zeta+-}
\end{equation}%
are shown to lead, in the limit $z\rightarrow 0$ (when both $z_{+}$\ and $%
z_{-}$\ reduce to $x$) and after some approximations, to the Schr\"{o}dinger
equation in terms of the parameter $\eta $ (to be determined below) 
\begin{equation}
-2\frac{\eta ^{2}}{m}\frac{\partial ^{2}\psi }{\partial x^{2}}+V(x)\psi
=2i\eta \frac{\partial \psi }{\partial t}  \label{Schrbeta}
\end{equation}%
and its complex conjugate, with 
\begin{equation}
\rho (x,t)=\psi ^{\ast }(x)\psi (x).  \label{rho}
\end{equation}%
It is important to stress that Eq. (\ref{rho}) is an integral part of the
theory; it is not a subsidiary postulate. This result indicates that a
regime of unitary (time-reversible) evolution has been attained, in which
the mechanical subsystem has acquired its quantum properties \cite%
{PeVaCeFr11}, \cite{CePeVa12}.

\subsection{The meaning of $\hbar $ in the Schr\"{o}dinger equation\label%
{balance}}

In order for (\ref{Schrbeta}) to be fully equivalent to the Schr\"{o}dinger
equation, the value of the parameter $\eta $ appearing in it must be
independent of the problem, and equal to $\hslash /2$. Although the
calculation of $\eta $ has been presented in previous work \cite{CePeVa12}, 
\cite{CePe12}, we briefly reproduce it here because it serves to disclose
the precise point of entry of Planck's constant into the Schr\"{o}dinger
equation.\footnote{%
In addition, the present calculation serves to correct a factor $1/2$
mistakenly introduced in previously published versions of Eq. (\ref{dHdt})
and the following.}

The value of $\eta $\ will be determined by resorting to the energy-balance
condition, which equates the average power lost by the particle through
Larmor radiation, to the average power extracted by the particle from the
random field. To establish the energy-balance condition we take the
generalized Fokker-Planck equation (\ref{FPE}), namely%
\begin{equation}
\frac{\partial }{\partial t}Q+\frac{1}{m}\frac{\partial }{\partial x}pQ+%
\frac{\partial }{\partial p}(f+m\tau \dddot{x})Q=e^{2}\frac{\partial }{%
\partial p}\hat{D}(t)Q,  \label{FPEcompl}
\end{equation}%
($p=m\dot{x}$) multiply it by $p^{2}$ and integrate over the entire particle
phase space. Assuming all surface terms to vanish at infinity, we obtain 
\begin{equation*}
\frac{1}{2m}\frac{d}{dt}\left\langle p^{2}\right\rangle =\frac{1}{2m}\frac{d%
}{dt}\int p^{2}Qdxdp=\frac{1}{m}\left\langle fp+m\tau p\dddot{x}-e^{2}p\hat{D%
}\right\rangle ,
\end{equation*}%
where $\left\langle g\right\rangle =\int g(x,p)Qdxdp.$ Since $d\left\langle
V\right\rangle /dt=-\left\langle fp\right\rangle /m$, the total average
energy gain or loss per unit time is given by%
\begin{equation}
\frac{d}{dt}\left\langle H\right\rangle =\frac{d}{dt}\left\langle \frac{1}{2m%
}p^{2}+V\right\rangle =m\tau \left\langle \dot{x}\,\dddot{x}\right\rangle -%
\frac{e^{2}}{m}\left\langle p\hat{D}\right\rangle ,  \label{dHdt}
\end{equation}%
where $H$ is the mechanical Hamiltonian function.{\tiny .} The first term on
the right-hand side represents the average power dissipated by the particle
through Larmor radiation; the second term represents the average power
extracted by the particle from the \textsc{zpf} and absorbed by the momentum
fluctuations. For energy balance to hold in the mean, these terms must
compensate each other, i.e., 
\begin{equation}
m\tau \left\langle \dot{x}\,\dddot{x}\right\rangle =\frac{e^{2}}{m}%
\left\langle p\hat{D}\right\rangle .  \label{EB}
\end{equation}

For a calculation of these two average values to lowest order in $\tau \sim
e^{2}$ we use the solutions of Eq. (\ref{Schrbeta}) containing the parameter 
$\eta $. Since the \textsc{zpf} represents the background field in its
ground state, the particle must also be in its ground state (denoted with
the subindex $0$). For the left-hand side of Eq. (\ref{EB}) this leads to 
\begin{equation}
m\tau \left\langle \dot{x}\,\dddot{x}\right\rangle _{0}=-m\tau
\dsum\limits_{k}\omega _{0k}^{4}\left\vert x_{0k}\right\vert ^{2},
\label{lhs}
\end{equation}%
where $\omega _{0k}=\left( \mathcal{E}_{0}-\mathcal{E}_{k}\right) /2\eta $, $%
x_{0k}=\int \psi _{0}^{\ast }x\psi _{k}dx$, $\mathcal{E}_{k}$ are\ the
energy eigenvalues and $\psi _{k}$ the corresponding eigenfunctions. For the
calculation of the right-hand side of Eq. (\ref{EB}), which is somewhat more
elaborate, we introduce Eq. (\ref{rho0}) for the spectral energy density of
the field ---which is proportional to $\hslash $--- into Eq. (\ref{diff}),
whence%
\begin{equation}
\frac{e^{2}}{m}\left\langle p\hat{D}\right\rangle _{0}=\frac{\hslash \tau }{%
\pi }\dint d\omega \,\omega ^{3}\dint dt^{\prime }\cos \omega (t-t^{\prime
})I(t-t^{\prime })  \label{int int}
\end{equation}%
with 
\begin{equation}
I(t-t^{\prime })=\dint dx\dint dp\,p\,e^{-\hat{L}(t-t^{\prime })}\frac{%
\partial }{\partial p}Q(t^{\prime })=\dint dx\dint dp\,p\,\frac{\partial }{%
\partial p^{\prime }}Q(x^{\prime },p^{\prime },t^{\prime }),  \label{Itt'}
\end{equation}%
where $x^{\prime },p^{\prime }$ are the position and momentum variables,
respectively, which evolve deterministically (under the action of $\hat{L}$)
towards their final values $x=x(t),$ $p=p(t)$. Upon integration by parts,
and writing $\dint dxdp=\dint dx^{\prime }dp^{\prime }$ to zero order in $%
e^{2}$, we get 
\begin{equation}
I(t-t^{\prime })=\left\langle \frac{\partial p}{\partial p^{\prime }}%
\right\rangle _{0}=\frac{1}{2i\eta }\left\langle [\hat{x}^{\prime },\hat{p}%
]\right\rangle _{0}=\frac{m}{\eta }\sum_{k}\omega _{k0}\left\vert
x_{0k}\right\vert ^{2}\cos \omega _{k0}(t-t^{\prime }).  \label{Itt'calc}
\end{equation}%
Now we insert (\ref{Itt'calc}) into (\ref{int int}) and integrate over time
starting at $-\infty $ (with $y=t-t^{\prime }$), to take into account that
energy balance is established after particle and field have interacted for a
sufficiently long time, i.e. in the so-called time-asymptotic limit 
\begin{eqnarray}
\frac{e^{2}}{m}\left\langle p\hat{D}\right\rangle _{0} &=&-\frac{\hslash
m\tau }{\pi \eta }\dsum\limits_{k}\omega _{k0}\left\vert x_{0k}\right\vert
^{2}\dint_{0}^{\infty }d\omega \ \omega ^{3}\int_{0}^{\infty }dy\cos \omega
y\cos \omega _{k0}y  \notag \\
&=&-\frac{\hslash m\tau }{2\eta }\dsum\limits_{k}\omega _{k0}\left\vert
x_{0k}\right\vert ^{2}\dint_{0}^{\infty }d\omega \ \omega ^{3}[\delta
(\omega +\omega _{k0})+\delta (\omega -\omega _{k0})].  \label{rhscalc}
\end{eqnarray}%
For $\omega _{k0}>0$ (as is the case for the ground state) the first
integral is nil, whence%
\begin{equation}
\frac{e^{2}}{m}\left\langle p\hat{D}\right\rangle _{0}=-\frac{\hslash m\tau 
}{2\eta }\sum_{k}\omega _{0k}^{4}\left\vert x_{0k}\right\vert ^{2}.
\label{rhs}
\end{equation}%
On comparing with Eq. (\ref{lhs}) we obtain 
\begin{equation}
\eta =\hslash /2,  \label{eta}
\end{equation}%
and Eq. (\ref{Schrbeta}) becomes the Schr\"{o}dinger equation,%
\begin{equation}
i\hbar \frac{\partial \psi }{\partial t}=-\frac{\hslash ^{2}}{2m}\nabla
^{2}\psi +V\psi .  \label{Schr}
\end{equation}%
Note that the \textsc{zpf} has played a crucial role in leading to this
result. Firstly, it is the source of the Planck constant in this equation,
through the spectral energy density given by Eq. (\ref{rho0}). Further, a
field with energy spectrum proportional to $\omega ^{3}$ (responsible for
the $\omega _{0k}^{4}$ factor in Eq. (\ref{rhs})) is the single one that
guarantees \emph{detailed} balance, by ensuring that Eqs. (\ref{lhs}) and (%
\ref{rhs}) have exactly the same structure. This means that energy balance
holds not only globally but term by term, or for each frequency. This
differs essentially from the result obtained for a \emph{classical} multiply
periodic system in equilibrium with a radiation field, in which case balance
is attained only if the field has a Rayleigh-Jeans spectrum, proportional to 
$\omega ^{2}$ \cite{VlHu77}.

\subsection{Detailed balance for an excited harmonic oscillator}

Let us now consider an atom in an excited state $n$, with the background
field still being in its ground state (the \textsc{zpf}). Then instead of (%
\ref{lhs}) we have 
\begin{equation}
m\tau \left\langle \dot{x}\,\dddot{x}\right\rangle _{n}=-m\tau
\sum_{k}\omega _{nk}^{4}\left\vert x_{nk}\right\vert ^{2},  \label{lhsn}
\end{equation}%
and instead of (\ref{rhs}) (with $\eta =\hslash /2$) we have%
\begin{equation}
\frac{e^{2}}{m}\left\langle p\hat{D}\right\rangle _{n}=-m\tau \sum_{k}\omega
_{nk}^{4}\left\vert x_{nk}\right\vert ^{2}\text{sign}(\omega _{kn}).
\label{rhsn}
\end{equation}%
Whilst in Eq. (\ref{lhsn}) all terms have the same sign, Eq. (\ref{rhsn})
contains now a mixture of positive and negative terms. As a result there is
a net average loss of energy 
\begin{equation}
\frac{d}{dt}\left\langle H\right\rangle _{n}\equiv \frac{dH_{n}}{dt}=-m\tau
\sum_{k}\omega _{nk}^{4}\left\vert x_{nk}\right\vert ^{2}(1-\text{sign}%
(\omega _{kn}))=-2m\tau \sum_{k<n}\omega _{nk}^{4}\left\vert
x_{nk}\right\vert ^{2}.  \label{dHdt-}
\end{equation}%
Hence, as was to be expected, there cannot be detailed balance between the 
\textsc{zpf} and the atom in an excited state; in the sole presence of the 
\textsc{zpf}, only the ground state ($n=0$) is in equilibrium.

Let us now assume that also the background field is in an excited state, and
inquire whether in this case there can be equilibrium between particle and
field. We write the spectral density of the excited field as 
\begin{equation*}
\rho (\omega )=\rho _{0}(\omega )\gamma (\omega ),\quad \gamma (\omega )\geq
1
\end{equation*}%
where the additional contribution 
\begin{equation}
\rho _{a}(\omega )=\rho -\rho _{0}=\rho _{0}(\gamma -1)\equiv \rho
_{0}\gamma _{a}  \label{rhoa}
\end{equation}%
can represent an excitation of the background field or an external field.
Now observe that the generalized form of Eq. (\ref{rhsn}) for the case $%
\gamma (\omega )>1,$ 
\begin{equation}
\frac{e^{2}}{m}\left\langle p\hat{D}\right\rangle _{n}=-m\tau \sum_{k}\omega
_{nk}^{4}\left\vert x_{nk}\right\vert ^{2}\gamma (\left\vert \omega
_{nk}\right\vert )\text{sign}(\omega _{kn}),  \label{rhsgamma}
\end{equation}%
contains again a mixture of terms with different signs depending on the sign
of $\omega _{kn}$, so that%
\begin{equation}
\frac{dH_{n}}{dt}=-m\tau \sum_{k}\omega _{nk}^{4}\left\vert
x_{nk}\right\vert ^{2}[1-\gamma (\left\vert \omega _{nk}\right\vert )\text{%
sign}(\omega _{kn})].  \label{dHdtgamma}
\end{equation}%
When the values of $\left\vert \omega _{nk}\right\vert $ differ for
different $k$, the positive and the negative terms in this equation cannot
compensate each other in general, and detailed balance can therefore not be
satisfied. However, if all values of $\left\vert \omega _{nk}\right\vert $
in Eq. (\ref{dHdtgamma}) are equal, it may be possible that detailed balance
exists with the particle in an excited state $n$. This is precisely the case
of the harmonic oscillator: all $\left\vert \omega _{nk}\right\vert $
appearing in (\ref{dHdtgamma}) are equal and moreover coincide with the
oscillator frequency $\omega _{0}$. With $\left\vert x_{nn+1}\right\vert
^{2}=a(n+1)$, $\left\vert x_{nn-1}\right\vert ^{2}=an,$ and $\left\vert
x_{nk}\right\vert ^{2}=0$ for $k\neq n\pm 1$, where $a=\hslash /2m\omega
_{0} $, Eqs. (\ref{lhsn}) and (\ref{rhsgamma}) give%
\begin{equation*}
m\tau \left\langle \dot{x}\,\dddot{x}\right\rangle _{n}=-\frac{1}{2}\hslash
\tau \omega _{0}^{3}(2n+1),\quad \frac{e^{2}}{2m}\left\langle p\hat{D}%
\right\rangle _{n}=-\frac{1}{2}\hslash \tau \omega _{0}^{3}\gamma (\omega
_{0}),
\end{equation*}%
whence the energy gain or loss is given by%
\begin{equation*}
\frac{dH_{n}}{dt}=-\frac{1}{2}\hslash \tau \omega _{0}^{3}[(2n+1)-\gamma
(\omega _{0})].
\end{equation*}%
Therefore, detailed balance exists between a harmonic oscillator in its
excited state $n$ and an (excited) background field with $\gamma (\omega
)=2n+1.$ According to Eq. (\ref{rhoa}), this field has a spectral energy
density 
\begin{equation}
\rho _{n}(\omega )=\rho _{0}(\omega )(2n+1)=\frac{\hbar \omega ^{3}}{2\pi
^{2}c^{3}}(2n+1),  \label{rhon}
\end{equation}%
corresponding to an energy per normal mode $\frac{1}{2}\hslash \omega (2n+1)$%
, equal to the energy of the mechanical oscillators with which it is in
equilibrium. This is simply the condition for balance between field and
matter oscillators of the same frequency ---and a result that links with the
Planck distribution in the case of thermal equilibrium (see \cite{PeVaCe08}
and section \ref{spont} below).

\section{Spontaneous and induced transitions}

\subsection{Radiative lifetimes\label{AB}}

Equation (\ref{dHdtgamma}) determines the average rate of energy loss or
gain by the mechanical system in an excited state $n$ due to (upward or
downward) transitions to states $k$ with $k>n$ and $k<n$, respectively. In
order to analyze this equation in detail it is convenient to write $\gamma
(\omega )=1+\gamma _{a}(\omega ),$ as follows from Eq. (\ref{rhoa}), and
separate the positive from the negative terms,%
\begin{align}
\frac{dH_{n}}{dt}& =-m\tau \sum_{k}\omega _{nk}^{4}\left\vert
x_{nk}\right\vert ^{2}\left[ 1-(1+\gamma _{a}(\left\vert \omega
_{nk}\right\vert ))\text{sign}(\omega _{kn})\right]  \notag \\
& =m\tau \sum_{k}\omega _{nk}^{4}\left\vert x_{nk}\right\vert ^{2}\left[
(\gamma _{a})_{\omega _{kn}>0}-(2+\gamma _{a})_{\omega _{kn}<0}\right] .
\label{emisabs}
\end{align}%
The first term within brackets in (\ref{emisabs}) represents the upward
transitions (absorptions) and the second one, the downward transitions
(emissions). It is clear that upward transitions can take place only when
there is an additional field $\gamma _{a}(\omega _{kn})$ from which the atom
may absorb the necessary energy; in other words, the atom does not
(`spontaneously') absorb energy from the \textsc{zpf}. This is an important
point that explains, for example, why optical detectors, including
photographic plates, are not activated by the vacuum. Emissions, on the
other hand, can be either `spontaneous' (in presence of just the \textsc{zpf}%
) or else stimulated by the additional field $\gamma _{a}(\omega _{kn})$,
according to the second term in (\ref{emisabs}). Notice in particular that
the ground state ($n=0$, $\omega _{kn}>0$) is absolutely stable against
spontaneous transitions.

Let us now relate the coefficients appearing in the various terms of Eq. (%
\ref{emisabs}), with the respective Einstein $A$ and $B$ coefficients. By
writing the rate of energy change in terms of the latter%
\begin{equation}
\frac{dH_{n}}{dt}=\sum_{k>n}\hslash \left\vert \omega _{nk}\right\vert [\rho
_{a}(\left\vert \omega _{nk}\right\vert )B_{kn}]-\sum_{k<n}\hslash
\left\vert \omega _{nk}\right\vert [A_{nk}+\rho _{a}(\left\vert \omega
_{nk}\right\vert )B_{nk}]  \label{dHn}
\end{equation}%
and comparing term by term with (\ref{emisabs}), we obtain for the
`spontaneous' emission coefficient 
\begin{equation}
A_{nk}=\frac{2\tau m}{\hslash }\left\vert \omega _{nk}\right\vert
^{3}\left\vert x_{nk}\right\vert ^{2}=\frac{4e^{2}\left\vert \omega
_{nk}\right\vert ^{3}}{3\hslash c^{3}}\left\vert x_{nk}\right\vert ^{2}
\label{Ank}
\end{equation}%
and for the stimulated transition coefficients%
\begin{equation}
B_{nk}=B_{kn}=\frac{m\tau \left\vert \omega _{nk}\right\vert ^{4}\left\vert
x_{nk}\right\vert ^{2}\gamma _{a}(\left\vert \omega _{nk}\right\vert )}{%
\hslash \left\vert \omega _{nk}\right\vert \rho _{a}(\left\vert \omega
_{nk}\right\vert )}=\frac{4\pi ^{2}e^{2}}{3\hslash ^{2}}\left\vert
x_{nk}\right\vert ^{2},  \label{Bnk}
\end{equation}%
in full agreement with the respective \textsc{qed} formulas \cite{Louisell73}%
,\cite{Milonni94}.$^{\text{ }}$

\subsection{Spontaneous decay and the zero-point field\label{spont}}

One can frequently find in the literature that all the spontaneous decay is
attributed to either the vacuum fluctuations or radiation reaction, more
often to the latter (see e.g. (\cite{Milonni94}, \cite{Davidov65}-\cite%
{DaDuCo82}). Let us look at this issue from the perspective of the present
theory.

From Eqs. (\ref{Ank}) and (\ref{Bnk}), the ratio of the $A$ to $B$
coefficients is%
\begin{equation}
\frac{A_{nk}}{B_{nk}}=\frac{\hbar \left\vert \omega _{nk}\right\vert ^{3}}{%
\pi ^{2}c^{3}}=2\rho _{0}(\left\vert \omega _{nk}\right\vert ).  \label{AtoB}
\end{equation}%
\ Incidentally, this relation and the equality of the coefficients $%
B_{nk}=B_{kn}$ were predicted by Einstein on the basis of his statistical
considerations (\cite{Einstein17}; see below).

Notice in particular the factor 2 in equation (\ref{AtoB}). Given the
definition of the coefficients, one could expect the ratio in this equation
to correspond exactly to the spectral density of the \textsc{zpf}, which
would mean a factor of 1. However, as follows from Eq. (\ref{emisabs}), one
should actually interpret the factor $2$ as $2=\left( 1+1\right) $. One of
these two equal contributions to spontaneous decay is due to the effect of
the fluctuations impressed on the particle by the \textsc{zpf}; the other
one is due to Larmor radiation. Their equality (with opposite signs) leads
to the exact balance of these two contributions when the particle is in its
ground state, thus guaranteeing its stability (cf. Eq. (\ref{dHdt-}) for $%
n=0 $).

A brief digression is in place regarding the point at which Einstein
introduced quantization in his 1917 paper \cite{Einstein17}, so as to arrive
at the Planck distribution. It is frequently argued that he did so through
the assumption of discrete atomic levels. However, some time after
Einstein's original work, Einstein and Ehrenfest \cite{EinsteinEhrenfest23}
showed that this was not the case, by redoing the calculations with a
continuous distribution of atomic levels. In line with the results presented
here and in previous work \cite{PeVaCe09}, \cite{CePeVa12} quantization
enters through the introduction of a source that includes the \textsc{zpf},
able to generate `spontaneous'\ transitions. This can be easily verified by
omitting in the calculation any of the three terms that lead to matter-field
equilibrium: stimulated absorptions and emissions, or spontaneous emissions.
The absence of the latter leads to absurd results, as happens also with the
omission of stimulated absorptions. The omission of the term related to
stimulated emissions leads to the expression for the blackbody law proposed
by Wien, which correctly approximates Planck's law at low temperatures, so
it already contains some quantum principle due to the presence of the term
associated with spontaneous emissions. All this can be easily seen in the
present context by focusing on just two states $n$ and $k,$ with $\mathcal{E}%
_{n}-$ $\mathcal{E}_{k}=\hslash \omega _{nk}>0$ and respective populations $%
N_{n},N_{k}$. When the system is in thermal equilibrium at temperature%
{\LARGE \ }$T,$ the relation ($k_{B}$ is Boltzmann's constant) 
\begin{equation*}
N_{k}/N_{n}=\exp (\mathcal{E}_{n}-\mathcal{E}_{k})/k_{B}T
\end{equation*}%
holds (disregarding inconsequential degeneracies). Since according to Eq. (%
\ref{emisabs}) the number of emissions is proportional to $N_{n}\gamma
_{a}(\omega _{nk})$ and the number of absorptions is proportional to $%
N_{k}[2+\gamma _{a}(\omega _{nk})],$ from the (detailed) balance condition $%
N_{n}\gamma _{a}=N_{k}(2+\gamma _{a})$ one obtains indeed Planck's law (for
the thermal field) as is well known since 1917 \cite{Einstein17},%
\begin{equation}
\gamma _{a}(\omega _{nk})=\frac{2}{\exp (\mathcal{E}_{n}-\mathcal{E}%
_{k})/k_{B}T-1}.  \label{ga}
\end{equation}

\section{Radiative corrections to the energy}

The determination of the Lamb shift has been one of the most frequently
studied problems in \textsc{sed} and has produced some successful results in
the past, though basically restricted to the linear-force problem. Early
related works are \cite{SokTumanov56}-\cite{Sa74}; additional references can
be seen in \cite{Pena83} and \cite{dice}. The theory of \textsc{sed} as
developed recently and used in this paper, has the advantage of being
applicable to nonlinear forces in general and to the atomic problem in
particular. Since this theory includes the radiative terms from the outset,
we can use it also to derive general formulas for the radiative energy
corrections. Here we present a full derivation of the nonrelativistic atomic
Lamb shift and associated effects within the present framework. The results
obtained are in line with the predictions deriving from \textsc{qed};
however, the procedure followed for their derivation offers a clear picture
of their physical meaning and allows a comparison with alternative
interpretations of the Lamb shift found in the literature.

\subsection{The Lamb shift \label{Lamb}}

To calculate the energy corrections we go back to the generalized
Fokker-Planck equation (\ref{FPEcompl}) and multiply it now by $xp$ before
integrating over the entire phase space; the result is (assuming again all
surface terms to vanish at infinity) \cite{Pena80}%
\begin{equation}
\frac{d}{dt}\left\langle xp\right\rangle =\frac{1}{m}\left\langle
p^{2}\right\rangle +\left\langle xf\right\rangle +m\tau \left\langle x\,%
\dddot{x}\right\rangle -e^{2}\left\langle x\hat{D}\right\rangle .
\label{dxpdt}
\end{equation}%
In the radiationless approximation, corresponding to quantum mechanics, the
last two terms are neglected and Eq. (\ref{dxpdt}) reduces to 
\begin{equation}
\frac{d}{dt}\left\langle xp\right\rangle =\frac{1}{m}\left\langle
p^{2}\right\rangle +\left\langle xf\right\rangle .  \label{dxpdt0}
\end{equation}%
Further, in a stationary state (denoted again by $n$) the time derivative of 
$\left\langle xp\right\rangle $ is zero and Eq. (\ref{dxpdt0}) reduces to
the virial theorem, 
\begin{equation}
\frac{1}{m}\left\langle p^{2}\right\rangle _{n}=2\left\langle T\right\rangle
_{n}=-\left\langle xf\right\rangle _{n},  \label{VT}
\end{equation}%
where $\left\langle T\right\rangle _{n}$ is the average kinetic energy.
Hence Eq. (\ref{dxpdt}) can be interpreted as a time-dependent version of
the virial theorem, with radiative corrections included. Observe that here
the average is taken not over time, but over the full particle phase space,
which is equivalent to an ensemble average. This is but an example of
application of the ergodic properties acquired by the quantum states as
discussed in detail in \cite{PeCe07}, \cite{PeVaCe08}.

In the stationary state, the two previously neglected terms can therefore be
taken as radiative corrections to the (kinetic) energy, 
\begin{equation}
\delta \mathcal{E}_{n}=\delta \left\langle T\right\rangle _{n}=-\frac{m\tau 
}{2}\left\langle x\,\dddot{x}\right\rangle _{n}+\frac{e^{2}}{2}\left\langle x%
\hat{D}\right\rangle _{n}.  \label{deltaT}
\end{equation}%
This is a general expression for the Lamb shift. Notice that the general
laws derived here ---such as (\ref{dHdt}) or (\ref{deltaT})--- are foreign
to quantum theory, where the notion of diffusion (and the related diffusion
operator) does not appear at all.

The right%
\'{}%
hand side of Eq. (\ref{deltaT}) will again be calculated to lowest order in $%
e^{2}$, which means calculating the two average values, $\left\langle x\,%
\dddot{x}\right\rangle _{n}$ and $\left\langle x\hat{D}\right\rangle _{n},$
to zero order in $\alpha =e^{2}/\hslash c$. For the first one we get%
\begin{equation*}
-\frac{m\tau }{2}\left\langle x\,\dddot{x}\right\rangle _{n}=\frac{\tau }{2}%
\left\langle \dot{x}\,f\right\rangle _{n}=\frac{\tau }{2}\frac{d}{dt}%
\left\langle T\right\rangle _{n}=0,
\end{equation*}%
which means that the Larmor radiation term does not contribute to the energy
shift in the mean, in a stationary state. The correction to the energy comes
exclusively from the fluctuations due to the action of the background field
on the particle, represented by the second term in Eq. (\ref{deltaT}). To
calculate it we use again Eq. (\ref{diff}), multiply it this time by $x$,%
\begin{equation*}
e^{2}x\hat{D}(t)Q=\frac{2\hslash }{3\pi c^{3}}e^{2}\dint d\omega \dint
dt^{\prime }\omega ^{3}\cos \omega (t-t^{\prime })\,x\,e^{-\hat{L}%
(t-t^{\prime })}\frac{\partial }{\partial p}Q(t^{\prime }),
\end{equation*}%
and integrate over phase space, thus obtaining%
\begin{equation}
\frac{e^{2}}{2}\left\langle x\hat{D}\right\rangle _{n}=\frac{\hslash e^{2}}{%
3\pi c^{3}}\dint d\omega \,\omega ^{3}\dint dt^{\prime }\cos \omega
(t-t^{\prime })J_{n}(t-t^{\prime }),  \label{xD}
\end{equation}%
with%
\begin{equation*}
J_{n}(t-t^{\prime })=\left. \dint dx\dint dp\,x\,e^{-\hat{L}(t-t^{\prime })}%
\frac{\partial }{\partial p}Q(t^{\prime })\right\vert _{n}.
\end{equation*}%
Upon an integration by parts (again with $\dint dxdp=\dint dx^{\prime
}dp^{\prime }$ to zero order in $\alpha $) we have $J_{n}(t-t^{\prime
})=-\left\langle \partial x/\partial p^{\prime }\right\rangle _{n},$ where 
\begin{equation}
J_{n}(t-t^{\prime })=\left\langle \frac{\partial x}{\partial p^{\prime }}%
\right\rangle _{n}=\frac{1}{i\hslash }\left\langle [\hat{x},\hat{x}^{\prime
}]\right\rangle _{n}=-\frac{2}{\hslash }\sum_{k}\left\vert x_{nk}\right\vert
^{2}\sin \omega _{kn}(t-t^{\prime }).  \notag
\end{equation}%
Inserting this result into (\ref{xD}) we obtain 
\begin{equation}
\frac{e^{2}}{2}\left\langle x\hat{D}\right\rangle _{n}=-\frac{2e^{2}}{3\pi
c^{3}}\dsum\limits_{k}\left\vert x_{nk}\right\vert ^{2}\dint_{0}^{\infty
}d\omega \ \omega ^{3}\int^{t}dt^{\prime }\cos \omega (t-t^{\prime })\sin
\omega _{kn}(t-t^{\prime }).  \label{xDcalc}
\end{equation}%
Extending the initial time integral to $-\infty $ (as corresponds to the
time-asymptotic limit) we have (with $y=t-t^{\prime }$)%
\begin{gather}
\int_{0}^{\infty }dy\cos \omega y\sin \omega _{kn}y=  \label{jnkt} \\
=\frac{1}{2}\int_{0}^{\infty }dy\left[ \sin (\omega _{kn}+\omega )y+\sin
(\omega _{kn}-\omega )y\right] =\frac{\omega _{kn}}{\omega _{kn}^{2}-\omega
^{2}},\text{ }  \notag
\end{gather}%
which introduced in Eq. (\ref{xDcalc}) gives for the radiative correction to
the (mean kinetic) energy (we write the result in three dimensions, for
comparison purposes){\LARGE \ }%
\begin{equation}
\delta \mathcal{E}_{n}=\frac{e^{2}}{2}\left\langle \mathbf{x\cdot \hat{D}}%
\right\rangle _{n}=-\frac{2e^{2}}{3\pi c^{3}}\dsum\limits_{k}\left\vert 
\mathbf{x}_{nk}\right\vert ^{2}\omega _{kn}\dint_{0}^{\infty }d\omega \ 
\frac{\omega ^{3}}{\omega _{kn}^{2}-\omega ^{2}}.  \label{deltaEn}
\end{equation}%
\qquad

This result coincides with the formula derived by Power \cite{Power66} for
the Lamb shift on the basis of Feynman's argument \cite{Feynman61}. We
recall that according to Feynman, the presence of the atom creates a weak
perturbation on the nearby field, thereby acting as a refracting medium. The
effect of this perturbation is to change the frequencies of the background
field from $\omega $ to $\omega /n(\omega )$, $n(\omega )$ being the
refractive index. The shift of the \textsc{zpf} energy due to the presence
of the atom in state $n$ is then (\cite{Milonni94}$,$\cite{Power66}) 
\begin{equation*}
\Delta \mathcal{E}_{n}=\dsum \frac{1}{2}\frac{\hslash \omega _{kn}}{%
n_{n}(\omega _{kn})}-\dsum \frac{1}{2}\hslash \omega _{kn}\simeq -\dsum
[n_{n}(\omega _{kn})-1]\frac{1}{2}\hslash \omega _{kn},
\end{equation*}%
where a summation over the polarizations is included, and the refractive
index is given in this approximation by{\LARGE \ }%
\begin{equation}
n_{n}(\omega )\simeq 1+\frac{4\pi }{3\hslash }\dsum\limits_{m}\frac{%
\left\vert \mathbf{d}_{mn}\right\vert ^{2}\omega _{mn}}{\omega
_{mn}^{2}-\omega ^{2}},  \label{nomegak}
\end{equation}%
where $\mathbf{d}_{mn}=e\mathbf{x}_{mn}$ is the transition dipole moment.
After an integration over the solid angle and summation over the
polarizations, Power obtains in the continuum limit for $\omega _{k}$ the
formula 
\begin{equation*}
\Delta \mathcal{E}_{n}=-\frac{2}{3\pi c^{3}}\dsum\limits_{m}\left\vert 
\mathbf{d}_{mn}\right\vert ^{2}\omega _{mn}\dint_{0}^{\infty }d\omega \ 
\frac{\omega ^{3}}{\omega _{mn}^{2}-\omega ^{2}},
\end{equation*}%
which is equal to our result, Eq. (\ref{deltaEn}).

The Lamb shift \textit{proper} (the observable Lamb shift) is obtained by
subtracting from the total energy shift given by Eq. (\ref{deltaEn}), the
free-particle contribution $\delta \mathcal{E}_{\text{fp}}.$ This latter is%
{\LARGE \ }represented by (\ref{deltaEn}) in the limit of continuous
electron energies (when $\omega _{kn}$ can be ignored compared with $\omega $
in the denominator),%
\begin{equation}
\delta \mathcal{E}_{\text{fp}}=\frac{2e^{2}}{3\pi c^{3}}\dsum\limits_{k}%
\left\vert \mathbf{x}_{nk}\right\vert ^{2}\omega _{kn}\dint_{0}^{\infty
}d\omega \ \omega =\frac{e^{2}\hslash }{\pi mc^{3}}\dint_{0}^{\infty
}d\omega \ \omega .  \label{Efp}
\end{equation}%
To write the last equality we used the sum rule $\Sigma _{k}\left\vert 
\mathbf{x}_{nk}\right\vert ^{2}\omega _{kn}=3\hslash /2m$. Since according
to Eqs. (\ref{corrE}) and (\ref{rho0}) 
\begin{equation}
\overline{\mathbf{A}^{2}}^{E}=\frac{2\hslash }{\pi c}\int_{0}^{\infty
}d\omega \ \omega ,
\end{equation}%
with $\mathbf{A}$ the electromagnetic potential associated with the \textsc{%
zpf}, (\ref{Efp}) can be rewritten as 
\begin{equation}
\delta \mathcal{E}_{\text{fp}}=\frac{1}{2m}\overline{\mathbf{A}^{2}}^{E},
\label{deltaEA2}
\end{equation}%
which identifies this `free-particle contribution' to the Lamb shift with
the contribution from the (free) background field. Its value is independent
of the state of the particle; it is a testimony of the ubiquitous presence
of the \textsc{zpf}. Inserting the usual cutoff frequency $\omega
_{c}=mc^{2}/\hslash $ in the integral in Eq. (\ref{Efp}) gives the finite
result $\delta \mathcal{E}_{\text{fp}}=(\alpha /2\pi )mc^{2}.$

By subtracting (\ref{Efp}) from (\ref{deltaEn}) we obtain for the Lamb shift
proper 
\begin{equation}
\delta \mathcal{E}_{\text{L}n}=\delta \mathcal{E}_{n}-\delta \mathcal{E}_{%
\text{fp}}=-\frac{2e^{2}}{3\pi c^{3}}\dsum\limits_{k}\left\vert \mathbf{x}%
_{nk}\right\vert ^{2}\omega _{kn}^{3}\dint_{0}^{\infty }d\omega \ \frac{%
\omega }{\omega _{kn}^{2}-\omega ^{2}}.  \label{ELamb}
\end{equation}%
Inserting once more the cutoff frequency $\omega _{c}=mc^{2}/\hslash $ in
the integral gives\footnote{%
Note that to get a correct (and finite) result, it is essential to leave in
the denominator of this formula the term $\tau ^{2}\omega _{mn}^{4}\simeq
\tau ^{2}\omega ^{4}$ under resonance due to the presence of radiation
reaction (\cite{SokTumanov56}, \cite{PeCe77}). This is a natural term in
both \textsc{qed} and \textsc{sed}.} 
\begin{equation}
\delta \mathcal{E}_{\text{L}n}=\frac{2e^{2}}{3\pi c^{3}}\dsum\limits_{k}%
\left\vert \mathbf{x}_{nk}\right\vert ^{2}\omega _{kn}^{3}\ln \left\vert 
\frac{mc^{2}}{\hslash \omega _{kn}}\right\vert ,
\end{equation}%
which is Bethe's well-known expression \cite{Bethe47}.

An important difference{\LARGE \ }between the procedures used in the present
paper and in \textsc{qed} to arrive at the Lamb shift formula concerns the
mass renormalization. We recall that in the \textsc{qed} case, second-order
perturbation theory is used, with the interaction Hamiltonian given by $\hat{%
H}_{\text{int}}=-(e/mc)\mathbf{\hat{A}\cdot \hat{p}.}$ But the energy
derived from this term, namely \cite{Milonni94}%
\begin{equation*}
-\frac{2e^{2}}{3\pi c^{3}}\dsum\limits_{k}\left\vert \mathbf{x}%
_{nk}\right\vert ^{2}\omega _{kn}^{2}\dint_{0}^{\infty }d\omega \ \frac{%
\omega }{\omega -\omega _{nk}},
\end{equation*}%
still contains the (linearly divergent) free-particle contribution%
\begin{equation*}
-\frac{2e^{2}}{3\pi c^{3}}\dsum\limits_{k}\left\vert \mathbf{x}%
_{nk}\right\vert ^{2}\omega _{kn}^{2}\dint_{0}^{\infty }d\omega =-\frac{%
4e^{2}}{3\pi c^{3}}\left( \frac{1}{2m}\dsum\limits_{k}\left\vert \mathbf{p}%
_{nk}\right\vert ^{2}\right) \dint_{0}^{\infty }d\omega
\end{equation*}%
that must be subtracted to obtain the Lamb shift proper. Because this result
is proportional to the mean kinetic energy, the ensuing correction is taken
to represent a mass renormalization,%
\begin{equation}
\delta m=\frac{4e^{2}}{3\pi c^{3}}\dint_{0}^{\infty }d\omega ,
\label{deltam}
\end{equation}%
which with the usual cutoff $\omega _{c}=mc^{2}/\hslash $ becomes $\delta
m=(4\alpha /3\pi )m$.

By contrast, in the derivation presented here to obtain $\delta \mathcal{E}_{%
\text{L}n},$ Eq. (\ref{ELamb}), there was no mass renormalization. The
result (\ref{deltam}) is just the \textit{classical} contribution to the
mass predicted by the Abraham-Lorentz equation (\cite{dice}, Eq. 3.114) (or
Maxwell's equations). In the equations of motion (\ref{eqsmot}) this
contribution has been already subtracted, so there is no more need to
renormalize the mass. However, and as is well known, the formula (\ref{ELamb}%
) (common to both \textsc{sed} and renormalized \textsc{qed}) still has a
logarithmic divergence that can be remedied by introducing the cutoff
frequency $\omega _{c}$, as was done by Bethe, thus obtaining a very
satisfactory result for the Lamb shift.

\subsection{Alternative interpretations of the Lamb shift}

The interpretation of the Lamb shift as a change of the atomic energy levels
due to the interaction with the surrounding \textsc{zpf} is fully in line
with the present theory. It constitutes one additional manifestation of the
influence of the particle on the field, which is then fed back on the
particle. An alternative way of looking at this reciprocal influence is by
considering the general relation between the atomic polarizability $\alpha $
and the refractive index of the medium affected by it, which for $n(\omega
)\simeq 1$ can be written as follows,%
\begin{equation*}
n(\omega )=1+2\pi \alpha (\omega ).
\end{equation*}%
Comparing this expression with Eq. (\ref{nomegak}) we obtain 
\begin{equation}
\alpha _{n}(\omega )=\frac{4\pi }{3\hslash }\dsum\limits_{m}\frac{\left\vert 
\mathbf{d}_{mn}\right\vert ^{2}\omega _{mn}}{\omega _{mn}^{2}-\omega ^{2}},
\label{alphan}
\end{equation}%
which is the Kramers-Heisenberg formula \cite{Davidov65}. This indicates
that the Lamb shift can also be viewed as a (second-order in $e\mathbf{E}$)
Stark shift associated with the dipole moment $\mathbf{d}(\omega )=\alpha
(\omega )\mathbf{E}$ induced by the electric component of the \textsc{zpf}
on the atom.

Equation (\ref{ELamb}) can be further recast in an (approximate) form that
is usual to find in textbooks and Lamb-shift calculations, by assuming that
the integral depends so weakly on the index $k$ that such dependence can be
ignored. Expressing (\ref{ELamb}) in terms of energy levels, with $\hbar
\omega _{nk}=\mathcal{E}_{n}-\mathcal{E}_{k},$ one gets then 
\begin{eqnarray*}
\delta \mathcal{E}_{\text{L}n} &=&-\frac{2e^{2}}{3\pi c^{3}}%
\dsum\limits_{k}\left\vert \mathbf{x}_{nk}\right\vert ^{2}\omega
_{kn}^{3}\dint_{0}^{\infty }d\mathcal{E}\ \frac{\mathcal{E}}{\left( \mathcal{%
E}_{k}-\mathcal{E}_{n}\right) ^{2}-\mathcal{E}^{2}} \\
&=&-\frac{2\alpha I_{n}}{3\pi c^{2}m^{2}}\dsum\limits_{k}\left\vert \mathbf{p%
}_{nk}\cdot \mathbf{p}_{kn}\right\vert ^{2}\left( \mathcal{E}_{k}-\mathcal{E}%
_{n}\right) ,
\end{eqnarray*}%
{\LARGE \ }which after a series of transformations becomes \cite{Milonni94}%
\begin{equation*}
\delta \mathcal{E}_{\text{L}n}=\frac{2\alpha I_{n}}{3\pi c^{2}m^{2}}i\hbar
\left\langle n\right\vert \nabla \hat{V}\cdot \mathbf{\hat{p}}\left\vert
n\right\rangle =\frac{\alpha \hbar ^{2}I_{n}}{3\pi c^{2}m^{2}}\left\langle
n\right\vert \left[ \nabla ^{2}\hat{V}\right] \left\vert n\right\rangle .
\end{equation*}%
This result suggests to interpret the Lamb shift as due to the variations of
the potential energy originating in the fluctuations in $\mathbf{x}$-space.
For the Coulomb potential, $\nabla ^{2}V=4\pi Ze^{2}\delta ^{3}(\mathbf{x}),$
so only the wave function at the origin ($s$ states) contributes to the Lamb
shift in this approximation. This makes this formula particularly practical
for numerical calculations.

\subsection{External effects on the radiative corrections}

From the results obtained above it is clear that certain basic properties of
the vacuum --- such as the intensity of its fluctuations or its spectral
distribution --- are directly reflected in the radiative corrections. This
means that a change in such properties can in principle lead to an
observable modification of these corrections. The background field can be
altered, for instance, by raising the temperature of the system, by adding
external radiation, or by introducing objects that affect the distribution
of the normal modes of the field.

Such external or `environmental' effects have been studied for over six
decades, normally within the framework of quantum theory, although some
calculations have been made also within \textsc{sed}, again for the
linear-force (or single-frequency) problem only, leading to comparable
results (see \cite{CePe881},\cite{CePe883}). The results of the previous
sections, by contrast, can be applied to the general case, without
restricting the calculations to the linear-force problem. In the following
we present an illustrative selection of results derived for both lifetimes
and energy levels. The task is facilitated by the use of the present theory
because the influence of the background radiation field (and its
modifications) is clearly pictured from the beginning.

In section \ref{AB} we have already come across one observable effect of a
change in the background field: according to Eq. (\ref{emisabs}) the rates
of stimulated atomic transitions are directly proportional to the spectral
distribution of the external (or additional) field, be it a thermal field or
otherwise. In the case of a thermal field in particular, with $\gamma
_{a}(\left\vert \omega _{nk}\right\vert )$ given by Eq. (\ref{ga}), the
(induced) transition rate from state $n$ to state $k$ becomes (using Eqs. (%
\ref{dHn}) and (\ref{Bnk})) 
\begin{equation}
\frac{dN_{nk}}{dt}=\rho _{0}(\left\vert \omega _{nk}\right\vert )\gamma
_{a}(\left\vert \omega _{nk}\right\vert )B_{nk}=\frac{4e^{2}\left\vert
\omega _{nk}\right\vert ^{3}\left\vert x_{nk}\right\vert ^{2}}{3\hslash c^{3}%
}\frac{1}{e^{\hslash \left\vert \omega _{nk}\right\vert /k_{B}T}-1}.
\label{dNn}
\end{equation}%
This result shows that no eigenstate is stable at $T>0$ --- as is well known
--- because the thermal field induces both upward and downward transitions.
For downward transitions ($\omega _{nk}>0)$ we can rewrite Eq. (\ref{dNn})
for comparison purposes in terms of $A_{nk}$ as given by Eq. (\ref{Ank}),
obtaining%
\begin{equation}
\frac{dN_{nk}}{dt}=\frac{A_{nk}}{e^{\hslash \left\vert \omega
_{nk}\right\vert /k_{B}T}-1}.
\end{equation}%
At room temperature ($k_{B}T\simeq .025$ eV) the effect of the thermal field
on the decay rate is barely noticeable, since for typical atomic frequencies
the inverse of the denominator, $\left( \exp \hslash \left\vert \omega
_{nk}\right\vert /k_{B}T-1\right) ^{-1},$ ranges between $\exp (-40)$ and $%
\exp (-400).$ The decay of excited states is therefore mostly spontaneous in
this case. For the thermal field to have a noticeable effect on the decay
rate, the temperature would have to be of the order of 10$^{4}$ K, at which
other effects on the atom (assuming it still exists at this high
temperature) cannot be ignored.

When the geometry or the spectral distribution of the field is modified by
the presence of conducting objects (such as metallic plates or the walls of
a cavity), the transition rates are affected accordingly. Let us assume, for
simplicity, that the modified field is still isotropic, with the density of
modes of a given frequency $\left\vert \omega _{nk}\right\vert $ reduced by
a factor $\gamma _{a}(\left\vert \omega _{nk}\right\vert )<1$: then
according to the results of section \ref{AB} the corresponding spontaneous
and induced transition rates are reduced by precisely this factor, since 
\textit{both} $A$ and $\rho B$ are proportional to the density of modes.%
\footnote{%
Interestingly, however, by virtue of this proportionality, the ratio of
spontaneous to induced transition rates is not altered by a modification of
the density of modes.} By enclosing the atoms in a high-quality cavity that
excludes the appropriate modes one can therefore virtually inhibit the
corresponding transition. For the more general (anisotropic) case the
calculations are somewhat more complicated, without however leading to a
substantial difference from a physical point of view. Such cavity effects
have been the subject of a large number of fine experimental tests since the
early works of Kleppner and others within \textsc{qed} (\cite{Kl81}-\cite%
{Martini87}; for more recent work see, e.g., \cite{Kreuter04},\cite%
{Walther06}).

For illustration purposes, let us also briefly indicate how Eqs. (\ref{Efp})
and (\ref{ELamb}) can be used to calculate the changes in the atomic energy
shifts produced by the addition of an (external or thermal) field. As in Eq.
(\ref{rhoa}), we denote by $\rho _{a}=\rho _{0}\gamma _{a}$ the spectral
(energy) density of the additional field. The formulas for the variations of
the (first-order) radiative corrections are readily obtained by determining
the shifts produced by the total field ($\rho _{0}+\rho _{a}$) and
subtracting the original shifts produced by the \textsc{zpf} alone. The
results are 
\begin{equation}
\Delta \left( \delta \mathcal{E}_{\text{fp}}\right) =\frac{e^{2}\hslash }{%
\pi mc^{3}}\dint_{0}^{\infty }d\omega \ \gamma _{a}\omega ,  \label{DeltaEfp}
\end{equation}%
\begin{equation}
\Delta \left( \delta \mathcal{E}_{\text{L}n}\right) =-\frac{2e^{2}}{3\pi
c^{3}}\dsum\limits_{k}\left\vert \mathbf{x}_{nk}\right\vert ^{2}\omega
_{kn}^{3}\dint_{0}^{\infty }d\omega \ \gamma _{a}\frac{\omega }{\omega
_{kn}^{2}-\omega ^{2}},  \label{DeltaELamb}
\end{equation}%
for a homogeneous, isotropic field. If the additional field represents
blackbody radiation at temperature $T$, $\gamma _{a}$ is given by Eq. (\ref%
{ga}), i.e. $\gamma _{a}(T)=2/(\exp y-1)$ with $y=(\hslash \omega /k_{B}T)$,
and we obtain from Eq. (\ref{DeltaEfp})%
\begin{equation}
\Delta _{T}\left( \delta \mathcal{E}_{\text{fp}}\right) =\frac{2\alpha }{\pi
mc^{2}}(k_{B}T)^{2}\dint_{0}^{\infty }dy\ \frac{y}{\exp y-1},
\label{DeltaTEfp}
\end{equation}%
whence the free-particle energy increases by the amount 
\begin{equation}
\Delta _{T}\left( \delta \mathcal{E}_{\text{fp}}\right) =\frac{\pi \alpha }{%
3mc^{2}}(k_{B}T)^{2}.  \label{DTEfp}
\end{equation}%
The formula for the change in the Lamb shift is given from Eq. (\ref%
{DeltaELamb}) by 
\begin{equation*}
\Delta \left( \delta \mathcal{E}_{\text{L}n}\right) =-\frac{4e^{2}}{3\pi
c^{3}}\dsum\limits_{k}\left\vert \mathbf{x}_{nk}\right\vert ^{2}\omega
_{kn}^{3}\dint_{0}^{\infty }d\omega \ \frac{\omega }{\omega _{kn}^{2}-\omega
^{2}}\left( \frac{1}{e^{\hslash \omega /kT}-1}\right) .
\end{equation*}%
These results coincide with those obtained through considerably more
cumbersome procedures within \textsc{qed} \cite{Knight72},\cite{ZhouYu10},
and the corresponding thermal shifts have also been experimentally confirmed
(\cite{HoHa84}; see also \cite{Brune94}). From the point of view of \textsc{%
sed} (or \textsc{qed}) their interpretation is clear: they represent
additional contributions to the kinetic energy impressed on the particle by
the thermal field, according to the discussion at the beginning of section %
\ref{Lamb}.

\section{Concluding remarks}

All results contained in this paper for a nonrelativistic spinless particle
point to the zero-point radiation field as the source not only of quantum
behavior itself, but also of the radiative effects on quantum systems. The
stochastic problem posed by the action of the \textsc{zpf} on the particle
led to a generalized Fokker-Planck equation for the particle, and all
results presented here have been directly derived from this equation.
Although as a result of the matter-\textsc{zpf} interaction both matter and
field end up quantized, for the present calculations of the radiative
corrections (to lowest order in $\alpha $) it sufficed to consider the 
\textsc{zpf} as a classical function. Indeed, the theory furnishes an
alternative way to derive self-consistently results usually considered to be
the exclusive province of \textsc{qed}. Important advantages of the present
procedure are physical transparency and simplicity; there was no need to
resort to heuristic arguments along the derivations. These advantages are
particularly apparent in the calculation of environmental effects on the
atomic lifetimes and energy levels.

It is important to stress that the present theory implies quantization of
both matter and radiation field \cite{PeVaCe09}, \cite{PeVaCeFr11}, \cite%
{PeVaCe08}. This is the ultimate reason that guarantees its equivalence with 
\textsc{qed}. The difference between these two theories lies not in their
final results, but in the whole conceptual picture and the gained
clarification of the physics. The present approach gives well-defined
answers to deep questions, such as the origin and ultimate meaning of the
Schr\"{o}dinger equation and other puzzles of quantum theory. The
consideration of the \textsc{zpf} as a fundamental ingredient of the theory
is thus not a subterfuge conceived to simplify or guide the calculations,
but a fundamental step to unfold the deep meaning of the quantum behaviour
of matter and field.

However, it should also be stressed that there can be differences. The
present theory gives, by construction, only an approximate description of
nature. It could be that its further development in search of a more
detailed or refined description leads to discrepancies, open to resolution
only by experiment.

Acknowledgment. This work was supported by DGAPA-UNAM through project PAPIIT
IN106412-2.

\end{document}